# NEXT-GENERATION HPC MODELS FOR FUTURE ROTORCRAFT APPLICATIONS

**NICOLETTA SANGUINI[1,\*], TOMMASO BENACCHIO[\*], DANIELE MALACRIDA[†], FEDERICO CIPOLLETTA[†], FRANCESCO RONDINA[†], ANTONIO SCIARAPPA[†] AND LUIGI CAPONE[†]**

[\*] Leonardo Labs, Future Rotorcraft Technologies, Leonardo S.p.A., Via G. Agusta 520, Samarate (VA), Italy {nicoletta.sanguini.ext, tommaso.benacchio.ext}@leonardo.com

[†] Leonardo Labs, Leonardo S.p.A., Via Pieragostini, 80, Genoa, Italy
{daniele.malacrida.ext, federico.cipolletta.ext, francesco.rondina.ext, antonio.sciarappa.ext, luigi.capone}@leonardo.com

**Abstract.** Rotorcraft technologies pose great scientific and industrial challenges for numerical computing. As available computational resources approach the exascale, finer scales and therefore more accurate simulations of engineering test cases become accessible. However, shifting legacy workflows and optimizing parallel efficiency and scalability of existing software on new hardware is often demanding. This paper reports preliminary results in CFD and structural dynamics simulations using the T106A Low Pressure Turbine (LPT) blade geometry on Leonardo S.p.A.'s davinci-1 high-performance computing (HPC) facility. Time to solution and scalability are assessed for commercial packages Ansys Fluent, STAR-CCM+, and ABAQUS, and the open-source scientific computing framework PyFR. In direct numerical simulations of compressible fluid flow, normalized time to solution values obtained using PyFR are found to be up to 8 times smaller than those obtained using Fluent and STAR-CCM+. The findings extend to the incompressible case. All models offer weak and strong scaling in tests performed on up to 48 compute nodes, each with 4 Nvidia A100 GPUs. In linear elasticity simulations with ABAQUS, both the iterative solver and the direct solver provide speedup in preliminary scaling tests, with the iterative solver outperforming the direct solver in terms of time-to-solution and memory usage. The results provide a first indication of the potential of HPC architectures in scaling engineering applications towards certification by simulation, and the first step for the Company towards the use of cutting-edge HPC toolkits in the field of Rotorcraft technologies.

## 1 INTRODUCTION

In the past twenty years, the exponential growth in computational power supplied by HPC clusters has made an increasingly broad range of scales accessible to computational fluid dynamics and structural mechanics simulations, along with great potential efficiencies in operational workflows.

---

[1] Corresponding author

In this context, aerodynamics of rotating-wing aircraft presents unique challenges to full-scale numerical representations. For external aerodynamics, the need to simulate high-Reynolds number, high-Mach number flows typically forces practitioners to use steady, averaged models, as time-varying direct numerical simulations on feature-resolving meshes imply times to solution that are currently incompatible with aircraft production schedules. However, emerging computational frameworks harnessing heterogeneous HPC systems have the potential to change that landscape and make scale-resolving simulations available [1].

This paper aims at evaluating the performance of the open-source framework PyFR and commercial packages Ansys Fluent and STAR-CCM+ in time-varying CFD simulations as well as of the ABAQUS framework in structural dynamics simulations on Leonardo S.p.A.'s HPC facility davinci-1. For the CFD simulations, particular attention is given to performance and scalability on GPU nodes. In this first assessment, we consider a relatively simple geometry, i.e., a single LPT blade T106A [2], and focus on a preliminary comparison of software tools in terms of fidelity of results, time to solution and scalability.

The work described in this paper was carried out at Leonardo S.p.A.'s Corporate Research Laboratory Leonardo Labs. Leonardo S.p.A., and Leonardo Helicopters Division in particular, routinely carry out CFD-based analysis and optimization for, e.g., rotorcraft design, layout and sizing, aerodynamic characterization, and validation of wind tunnel data, helping reduce industrial costs associated with physical prototypes and wind tunnel testing.

The paper is organized as follows. Section 2 provides details on the software and hardware setup for the CFD and structural simulations. Section 3 reports details on the mesh used, the test case setup, and numerical results on fluid flow simulations and structural simulations, and Section 4 contains final considerations and an outlook to future work.

## 2 COMPUTATIONAL SETUP

### 2.1 Software packages

This paper considers several numerical frameworks for computational fluid dynamics and structural dynamics simulations. For computational fluid dynamics simulations, commercial packages Ansys Fluent[2] and Simcenter STAR-CCM+[3] were chosen as they are widely used tools in an industrial context, including at Leonardo S.p.A. They offer a complete workflow of pre-processing tools, computational models for compressible and incompressible fluid flow simulations, and post-processing capabilities with interactive GUIs as well as batch running modes. While Fluent already has support for running on GPU-based architectures, STAR-CCM+ only has GPU support for incompressible runs. Both packages employ finite-volume based numerical methods that are at most second-order accurate. In addition, the open-source framework PyFR was considered for CFD runs. PyFR [3] solves the compressible fluid flow equations using the element-based flux reconstruction method [4] without the use of turbulence models, has the option of running at high order via the use of high-degree

---

[2] https://www.ansys.com/products/fluids/ansys-fluent
[3] https://www.plm.automation.siemens.com/global/en/products/simcenter/STAR-CCM.html

N. Sanguini, T. Benacchio, D. Malacrida, F. Cipolletta, F. Rondina, A. Sciarappa, and L. Capone

polynomials, and accommodates support for unstructured meshes with several element shapes, thereby lending itself well to simulations on complex geometries. Thanks to its data locality properties - higher accuracy is achieved at higher polynomial order using multiple degrees of freedom inside a single mesh element - the flux reconstruction method has advantageous parallelization capabilities compared to wide stencils-based higher-order extensions of finite-volume methods. In terms of implementation, PyFR's Python code base with runtime-metaprogramming and embedded domain-specific language technology was designed for hardware portability. In particular, PyFR has native GPU support via runtime-generated OpenCL, CUDA, and HIP backends, through which the whole code base is automatically optimized for high performance on GPU-based HPC clusters. The code was successfully tested on industrial LPT blade simulations [1], showcased excellent scalability and efficiency on thousands of GPUs [5], and was previously compared with STAR-CCM+ [6]. Inset 1, inspired by [6], provides details of the CFD software versions used in this paper. In order to provide as fair a comparison as possible, the times-to-solution and scalability figures for Fluent and STAR-CCM+ refer to runs in Laminar mode, i.e. without the use of any turbulence model (see also details on parameter choices in Section 3.1 and discussion in Section 4).

For the structural mechanics simulations, the commercial software ABAQUS[4] was chosen for its wide adoption in industrial workflows. The software offers a unified environment to generate both static and dynamic structural cases on a wide range of finite element formulations, supporting several load cases and interaction algorithms to model complex full-scale problems consisting of several hundreds of parts. In addition, its robust solvers can be fine-tuned by the user in order to solve nonlinearities, with minimal impact on computational resources.

| **Functionality summary of PyFR v. 1.12.2** | |
|---|---|
| Systems | Compressible Euler, Navier-Stokes |
| Element Types | Triangles, quads, hexahedra, prisms, tetrahedra, pyramids |
| Platforms | CPU, GPU (NVIDIA and AMD) |
| Spatial Discretization | Flux reconstruction (choice of polynomial order p) |
| Temporal discretization | Explicit |
| Turbulence model | None (DNS) |
| **Functionality summary of Ansys Fluent v2021R1** | |
| Systems | Compressible Euler, Navier-Stokes |
| Element Types | Tetrahedral, polyhedral, triangles, quads |
| Platforms | CPU, GPU |
| Spatial Discretization | Finite volume |
| Temporal discretization | Explicit |
| Turbulence model | Laminar or LES |
| **Functionality summary of SimCenter STAR-CCM+ v2022.1** | |
| Systems | Compressible Euler, Navier-Stokes, etc. |
| Element Types | Tetrahedral, polyhedral, etc. |
| Platforms | CPU (GPU incompressible only) |
| Spatial Discretization | Finite volume |
| Temporal discretization | Explicit, implicit |
| Turbulence model | Laminar or LES |

**Inset 1.** CFD computational frameworks considered in this paper.

---

[4] https://www.3ds.com/products-services/simulia/products/abaqus/

## 2.2 Hardware infrastructure: davinci-1

CFD and structural mechanics simulations in this paper were performed using Leonardo S.p.A.'s HPC facility davinci-1, which has been operative at Leonardo since 2020 to support Corporate R&D activities. Davinci-1 has 116 computing nodes with 48 CPU cores each. A partition of 80 computing nodes is accelerated with 4 NVidia A100 GPUs per node. Davinci-1 supplies a peak performance of 6.5 Petaflops, ranking #150 overall, and #4 in the AD&S sector, on the Top500 list (June 2022 standings)[5]. Due to license issues, only 10 computing nodes could be used for runs with Fluent. Figure 1 contains more details on the davinci-1 infrastructure.

**Compute nodes (CPU) x 56:**
- 2x24 cores Intel(R) Xeon(R) Platinum 8260 CPU @ 2.40GHz
- 1024 GB total DRAM memory
- 456.32 GB/s factory memory bandwidth per node

**Compute nodes (GPU) x 60 + (FAT) x 20:**
- 2x24 cores AMD EPYC Rome 7402 CPU @ 2.80GHz
- 512 GB total DRAM memory (FAT: 1024GB)
- 4 x A100-SXM4-40GB Nvidia GPUs
- 9.746 TFLOPS in FP64 performance
- 663.6 GB/s factory memory bandwidth per node

**Figure 1.** davinci-1 computational resources.

## 3 NUMERICAL RESULTS

### 3.1 Flow around an LPT blade

In this study, we focus on the T106A LPT blade geometry[6] [2]. The simulation setup has a single blade exposed to uniform inflow. To consider a slightly different dynamics from what already seen in the literature, a cylinder was placed in front of the blade as a wake generator.

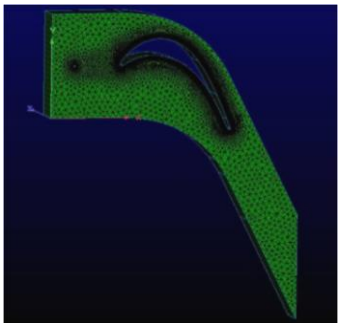

**Figure 2.** Unstructured mesh for the CFD experiments.

The geometry of the blade was normalized with respect to the chord $c$ and the span $h$ set as $h = c = 1$. The mesh was generated with the Pointwise[7] software by extrusion starting from the two-dimensional triangle-based unstructured mesh on the side wall. To solve the boundary layers near the surface of the blade and of the cylinder, we used the T-Rex method within Pointwise and generated a mesh with 50 layers of quadrilaterals below the triangles. The minimum wall spacing was set at $6e^{-5}$ and a growth rate of 1.1 was used (Figure 2).

Along the spanwise direction, $N = 40$ elements of length $l = h/N$ were created by extrusion. The resulting three-dimensional mesh has 663,240 elements (483,400 hexahedra and 179,840 prisms). For some tests in this work, additional meshes were built by modifying the length of the span $h$, and consequently the number of elements $N$, while keeping the length of the cells $l$ constant. For the initial conditions, the velocity is set to zero. The boundary conditions prescribe inlet and outlet velocity, and periodic upper, lower, and lateral walls.

---

[5] https://www.top500.org
[6] http://www-g.eng.cam.ac.uk/whittle/T106/Start.html
[7] https://www.pointwise.com

In the first test, we performed the simulation with PyFR at different polynomial order $p$. In particular, we considered $p = 1$ and $p = 2$. To set the time step, we conservatively imposed an initial Courant number equal to 0.3. As a result, for $p = 1$ the resulting initial time step was $1.5e^{-5}$, while for $p = 2$ the time stamp was $1e^{-5}$. We consider dimensionless dynamic viscosity $\mu = 2{,}77e^{-6}$. For the compressible simulation, a purely horizontal velocity with dimensionless magnitude 0.114 was set at the inlet. As in [2], at the outflow an angle of $-63.2$ and a velocity magnitude of 0.495 are imposed. The inlet flow was set along the streamwise direction, so that the flow arrives perturbed at the blade after passing through the wake generator. Based on the chord $c$, and considering a velocity value of 0.5 (see below) a Reynolds number of ~250,000 is obtained.

Figure 3 shows the velocity magnitude at $t = 100$ for the two simulations. In both cases, the numerical solution provides a stable representation of the turbulent dynamics in the wake of the cylinder and the blade. The maximum dimensionless velocity recorded is for both cases 0.53 (about $170 \ m/s$, maximum Mach number of around 0.5).

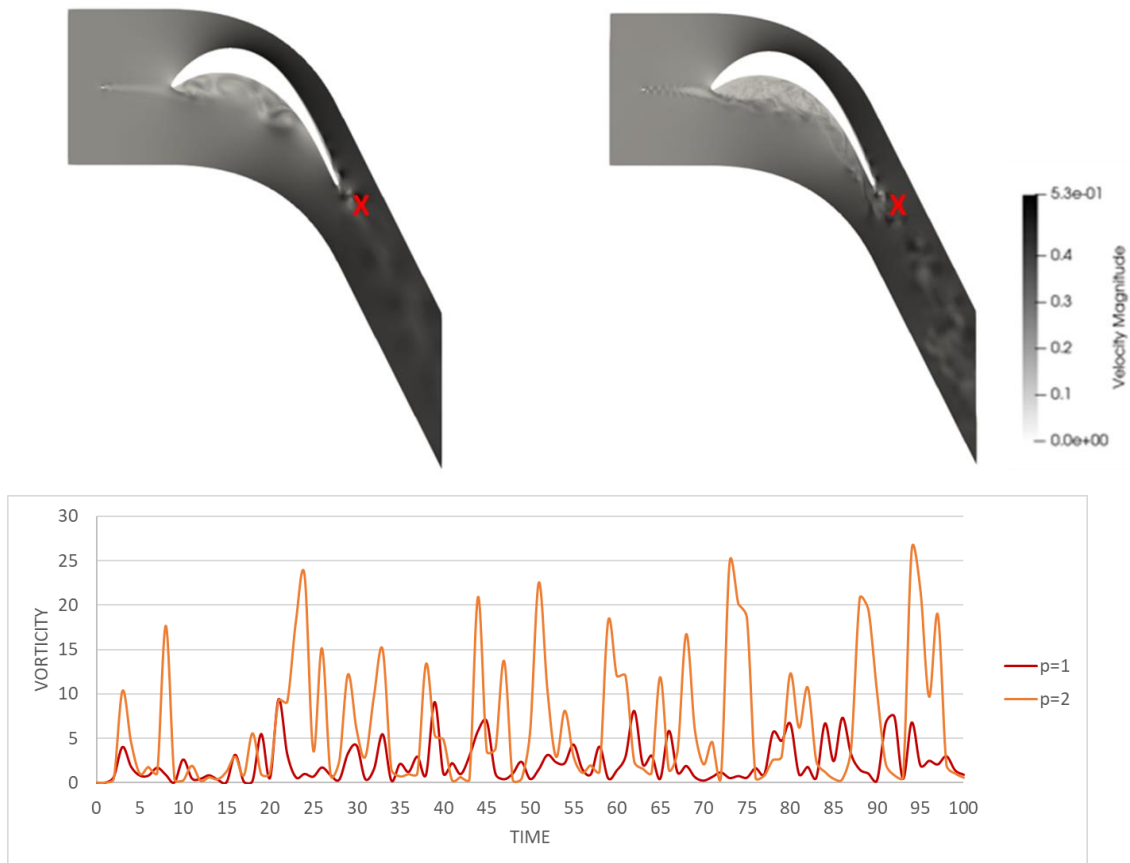

**Figure 3.** PyFR simulation on the 663.24K mesh. Top: velocity magnitude at $T = 0.06 \ s$ using polynomial degrees $p = 1$ (left) and $p = 2$ (right). Bottom: time series of vorticity magnitude in the wake of the blade (markers in top plots) using polynomial degrees $p = 1$ (red) and $p = 2$ (orange).

In addition, the higher-order solution with $p = 2$ is more detailed than the simulation with $p = 1$, especially in the wake of the blade. This is confirmed by a probe measurement for the vorticity magnitude at a point of the wake of the turbine blade (red crosses in the top of Figure 3). The time series (Figure 3 bottom) clearly reveals that, while the behavior is not perfectly periodic due to the wake being underresolved at both orders, the higher order simulation yields deeper vortices. This is to be expected in view of the higher order of accuracy and implied effective resolution - the considered mesh has about $4.97M$ degrees of freedom (DoFs) at $p = 1$, and two and a half times as many DoFs ($\sim 12.4M$) at $p = 2$. Preliminary results on a higher-order run at $p = 4$ (not shown) display deeper vortices than at lower orders.

In the second test, we compared the performance of PyFR and the two commercial packages Ansys Fluent and Star-CCM+. In this case, we considered a mesh with about $2.65M$ elements (i.e., with a $4h$ span). For all three simulations we used the same number of computational resources, i.e., 4 davinci-1 GPUs nodes. As regards Ansys Fluent and STAR-CCM+, compressible flow was considered in the laminar configuration (no subgrid-scale turbulence model). In both cases an unsteady implicit solver was used, with a time-step of $5e^{-6}\ s$. The dynamic viscosity was considered constant and equal to $1.789e^{-5}\ Pa*s$ to have the same Reynolds number of dimensionless PyFR simulations.

The results obtained by the three tools are qualitatively similar (Figure 4). StarCCM+ has a lower velocity than the other two tools, while the solution with PyFR contains more details of the dynamics than the two commercial codes. This is to be expected since Fluent and Star use finite volumes (1 DoF per element), while PyFR run with polynomial order p=1 (~7.5 DoF per element on the considered mesh).

In terms of time to solution, the wallclock time values of PyFR ($p = 1$) and Fluent are very similar, while Star is about 40% faster (Figure 5 left). This result is surprising, considering the fact that Star runs on CPUs without accelerators for this compressible case. However, the situation is quite different when considering the wallclock time per DoF (Figure 5 right). Here, PyFR is more than eight times faster than Fluent more than four times faster than Star.

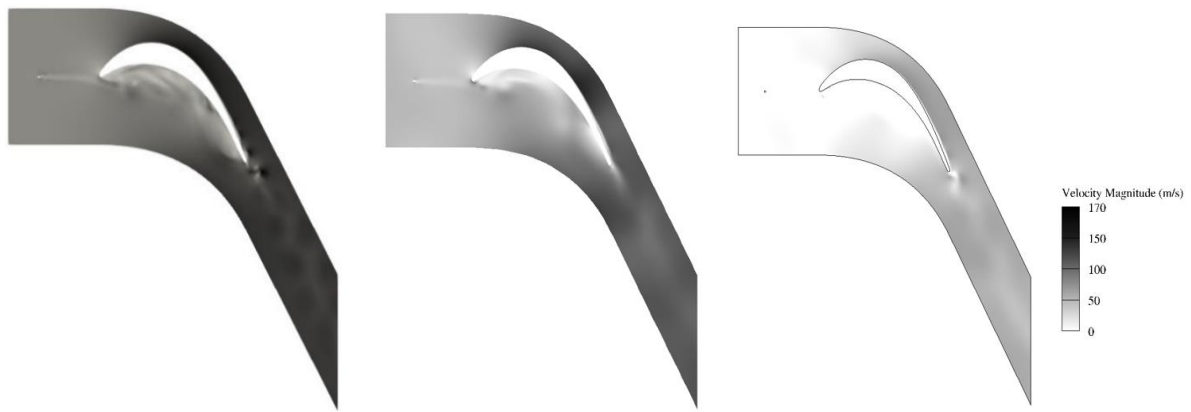

**Figure 4.** Velocity magnitude at $T = 0.02\ s$ in the T106A simulation with PyFR (p1, left), Fluent (center), and STAR-CCM+ (right), 2.65M elements mesh.

For further comparison, we also ran a simulation with PyFR on a coarser mesh ($\sim 180K$ elements), but at polynomial order $p = 4$, to have the same DoFs of the $p = 1$ case ($\sim 19.87M$ DoFs on the 2.65M element mesh). The total wallclock time in this case is equal to four times that of $p = 1$, but in terms of wallclock time per DoF, the result is still superior to both Fluent and Star, and at even higher accuracy.

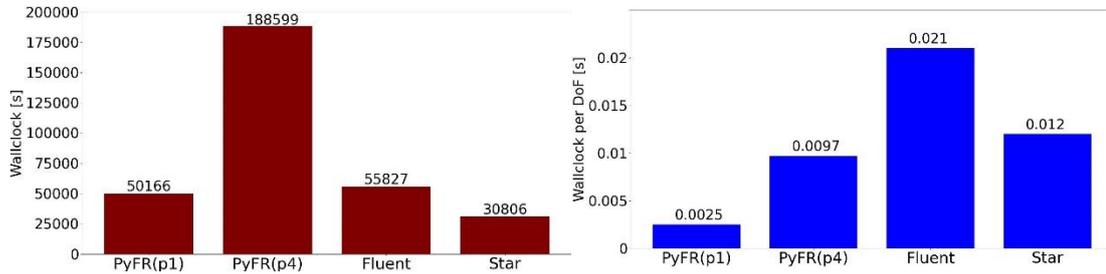

**Figure 5.** Wallclock time (left) and wallclock time / DoF (right) with PyFR ($p = 1$ and $p = 4$), Ansys Fluent, and STAR-CCM+ in the T106A simulation, $T = 0.02\ s$, 2.65M elements mesh.

Finally, we perform scalability tests on the GPUs. For the strong scaling of PyFR we use the same mesh with $2.65M$ elements as in the previous test and polynomial order $p = 1$. The simulations run on 2, 4, 8, 16 and 32 GPU nodes, each node using 4 GPUs. Figure 6 shows the speedup, on the top left, and the wallclock time (red bars) and the number of elements per GPUs (green line), on the top right.

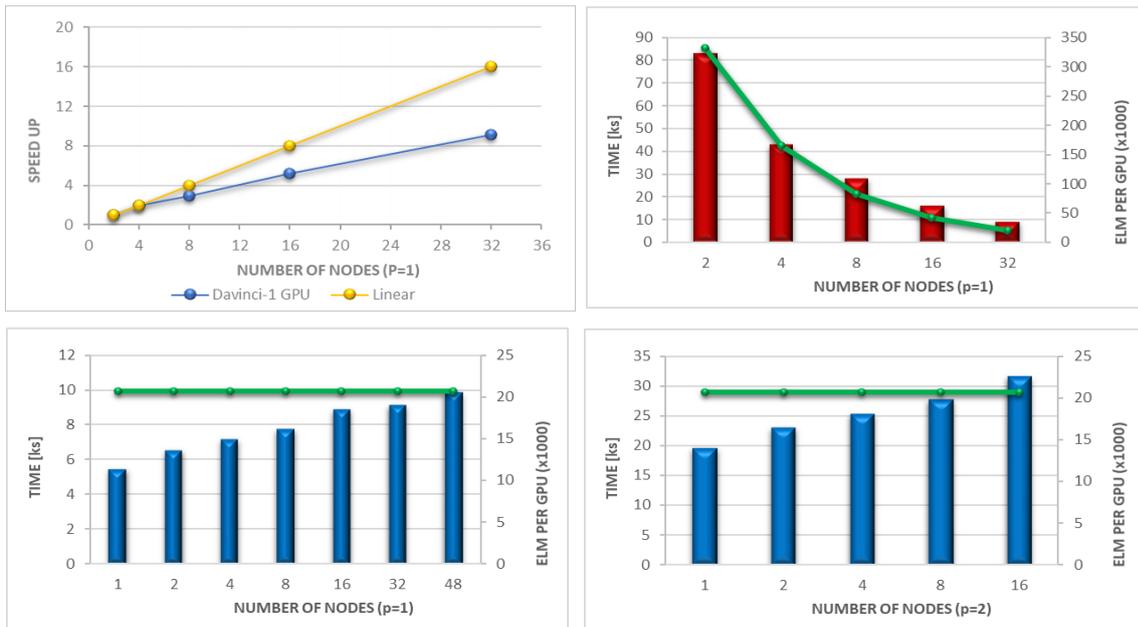

**Figure 6:** PyFR scaling. Top row: strong scaling at p=1, 2.65M elements mesh. Speedup (left) and wallclock time (right) on 2,4,8,16, and 32 davinci-1 nodes, each using 4 A100 GPUs. Bottom row: weak scaling at $p = 1$ (left, 1 to 48 nodes, 83K elements to 3.979M elements meshes), $p = 2$ (right, 1 to 16 nodes, 83K elements to 1.328M elements meshes). Here and in the following, the green line denotes the number of elements per GPU.

The obtained speedup is acceptable, considering that the code is run at relatively low order and without fine tuning. For the weak scaling experiment, the computational load is fixed at $20.7K$ elm/GPU, increasing the number of computing nodes proportionally to the problem size. Figure 6 shows on the bottom the results of the weak scaling at polynomial order $p = 1$ (left) and $p = 2$ (right). In particular, $p = 1$ runs from 1 node ($83K$ elements mesh, span = $h/8$) up to 48 nodes ($3.979M$ elements mesh, span = $6h$), while $p = 2$ runs from 1 node up to 16 nodes ($1.328M$ elements mesh, span = $2h$). In both cases the weak scaling results are not optimal, because the time to solution increases with the number of nodes. This result could be due to the low computational load per GPU. Better results should be obtained for scaling using higher polynomial orders on the same mesh.

For Fluent scalability, we used the best configuration tested with other benchmarks on davinci-1, i.e., full node configuration (48 CPU cores/node, 4 GPUs/node). For strong scaling we used a mesh with $21.2M$ elements (about 10 times larger than PyFR scaling but with comparable DoFs count at $p = 1$). The scaling test is run on 1 node to 9 nodes. The obtained speedup is suboptimal compared to the ideal linear trend, and in particular for 8 to 9 nodes the model antiscales (Figure 7 left). For weak scaling, we ran on 1, 2, 4 and 8 GPUs nodes, fixing the computational load at $663.24K$ elm/GPU (Figure 7 right). For the first three simulations the time difference is relatively small, while a more marked difference is noted with 8 nodes. We remark that we could not perform the scalability tests on the GPUs for this compressible case with Star-CCM+, which only supports GPU acceleration for incompressible flows.

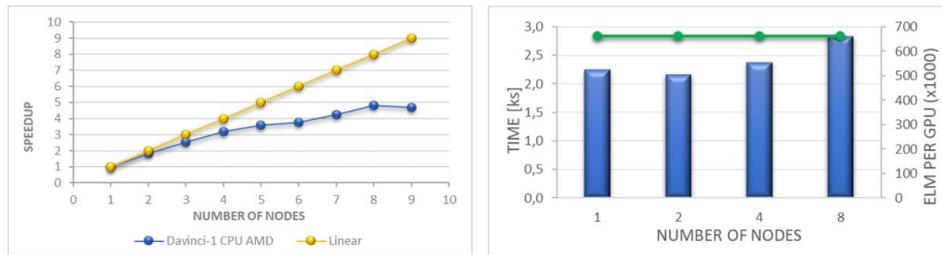

**Figure 7:** Ansys Fluent scaling results. Left: strong scaling, 2.65M elements mesh, speedup on 1 to 9 nodes, each using 4 A100 GPUs. Right: weak scaling, 1,2,4,8 nodes, 83K, 166K, 332K, 664K elements meshes.

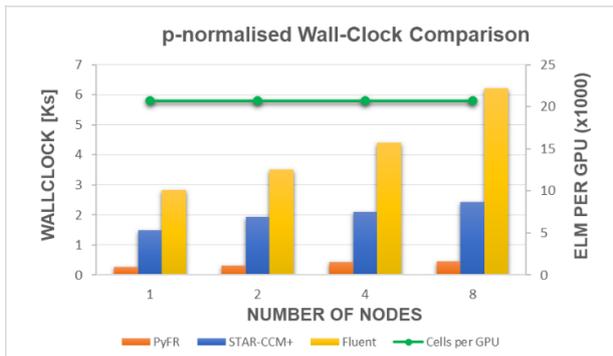

**Figure 8:** Weak scaling in the incompressible flow simulations with PyFR ($p = 1$, orange), STAR-CCM+ (blue), and Ansys Fluent (yellow), 2.65M elements mesh, on 1, 2, 4, 8 nodes, each running with 4 A100 GPUs. The PyFR wallclock time is normalized for the

To compare the performance of all three tools, we therefore ran a case in the incompressible regime, using a 10 times smaller inflow and outflow velocity values. In this way we obtain a Mach number lower than 0.1. In this case, the load per node is fixed at $20.7K$ elm/GPU, as in the previous PyFR weak scaling, and the simulations run on 1,2,4 and 8 GPU nodes. 8 shows the comparison of the weak scaling results performed by the three tools. The values in the bar graph are relative to the normalized wallclock time with respect to the number of DoFs (PyFR is set at $p = 1$). StarCCM+ and PyFR scale better than Fluent, and PyFR runs about 5 times faster than Star, and about 10 times faster than Fluent.

## 3.2 Linear elasticity simulations

Elasticity problems in engineering applications can contain several hundreds of physical components and include multiple features such as rigid bodies, connectors, tied links, in order to simulate connections between parts, like bolts, rivets, welding. Typical loads and boundary conditions take the form of distributed pressure and concentrated force or torque as well as imposed or prescribed displacements and are typically applied at mesh nodes.

Nonlinearities in the form of geometrical formulation, constitutive models, and contact formulations add complexity to the algebraic problem and require incremental solution schemes, where the convergence times becomes significantly problem dependent.

Traditionally, structural finite element codes are equipped with an extensive element library, in order to discretize geometries efficiently, avoid redundancies and minimize overhead in the analysis of specific problems. Nevertheless, the advent of HPC infrastructures offers the possibility to forego several constrains and allows the analysts to solve high-fidelity models spanning multiple length scales, opening the doors to new and advanced interactions and constitutive material formulations. Thanks to this, structural models containing several millions of DoFs have become commonplace, therefore, optimizing the usage of the available computational resources becomes a necessary step.

In order to provide a fair basis for optimization and remove the influence of nonlinearities, a simple case of static linear elasticity is investigated in detail on the davinci-1 HPC infrastructure using the ABAQUS framework.

The three-dimensional geometry of the T106A LPT blade is discretized with reduced integration hexahedral continuum elements (C3D8R), for a total of $1.58M$ elements and $5.0M$ DoFs. The part is loaded with distributed pressure around the profile, the root is constrained in all directions with encastre boundary conditions while the opposite end is left free (Figure 9).

A comparison of direct and iterative solvers under several resource configurations is performed. Results up to 48 cores refer to a single CPU compute node, while data points for the 96 cores are relative to a configuration using two nodes. The solution clock time and memory usage are extracted from the output logs generated by ABAQUS. The absolute solution time, memory usage and speedup against the single core solution are shown in Figure 10.

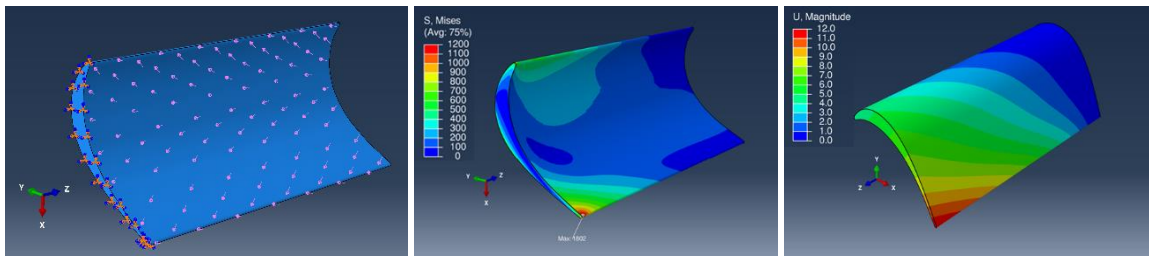

**Figure 9:** Linear elasticity simulations with the T106A geometry:applied loads and boundary conditions (left), von Mises stress contour plot (center), displacement magnitude (right).

Regarding the comparison between direct and iterative solver, it can be seen from Figure 10 (top left) that the iterative solver has an evident advantage in large problems like the one investigated. Furthermore, the memory usage for the iterative solver is significantly lower at 44GB compared to 97GB for the direct solver in single core solution. This gap widens when multiple resources are requested, as the direct solver shows a significant overhead, up to +160% when 96 MPI processes are requested, whereas memory overhead is negligible using the iterative solver (Figure 10, top right).

More specifically, along with the number of cores, declared in the *ncpus* parameter, ABAQUS allows the control of the number of MPI processes used for a solution, through the parameter *threads_per_mpi_process*, such that the number of MPI processes is determined as: #*ncpus* / #*threads_per_mpi_process*.

This setting is found to significantly affect the solution time, particularly for the iterative solver, which benefits vastly from a higher parallelization of the solution. The effect is less significant for the direct solver, where the speedup at different values of *threads_per_mpi_process* is similar at a given #*ncpus*. Nevertheless, the speedup obtained from the direct solver seems to hit a limit when the total number of processes is higher than 12.

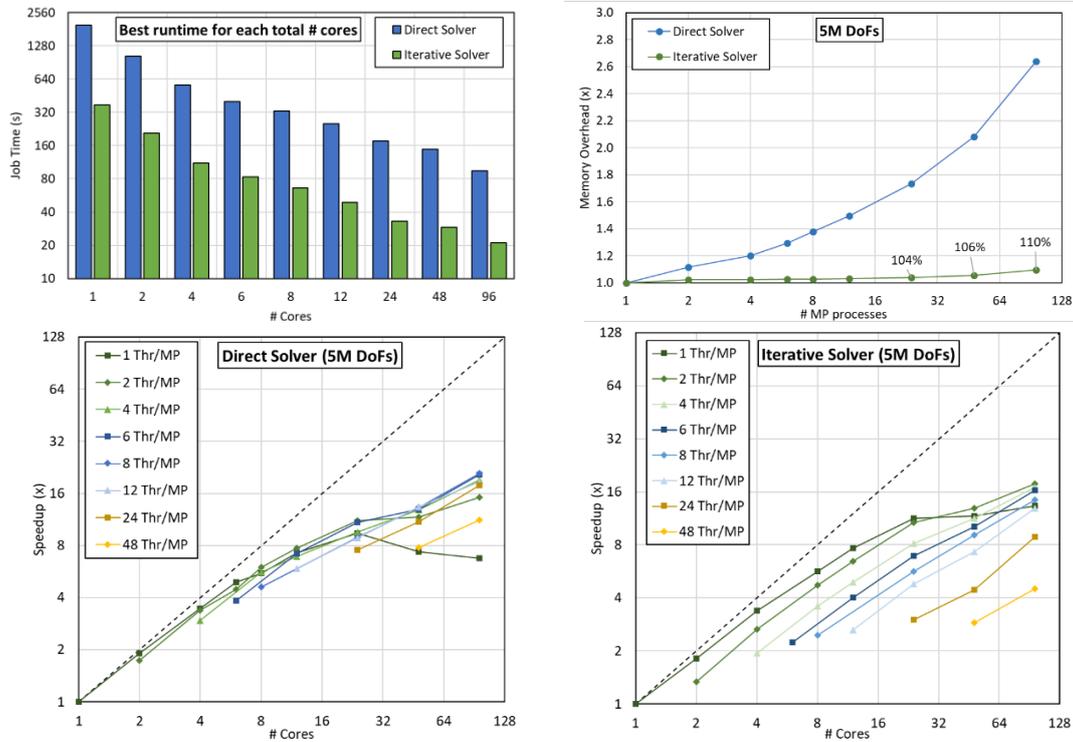

**Figure 10:** Top: strong scaling (left) and memory overhead (right) on davinci-1 CPUs in the T106A linear elasticity runs with ABAQUS. The 96 cores result uses 2 CPU nodes. Bottom: speedup at different settings of ncpus and threads_per_mpi_process for direct solver (left) and iterative solver (right).

## 4 CONCLUSIONS

This paper has investigated the performance of a heterogeneous HPC architecture, Leonardo S.p.A.'s davinci-1, on CFD and structural dynamics simulations using an idealized T106A LPT blade geometry. In simulations of compressible and incompressible flow around a wake generator and the LPT blade, the commercial packages Ansys Fluent and Simcenter STAR-CCM+ as well as the open-source high-order framework PyFR provided stable and scalable numerical solutions using hardware accelerators. Ansys Fluent provided strong and weak scaling up to 9 computing nodes with GPUs and up to 2.4 million elements per node (testing with more resources was constrained by license limitations). PyFR provided speedup up to 128 GPUs (down to about 20 thousand elements per GPU), and weak scaling up to 192 GPUs (i.e., 48 GPU nodes, 60% of the whole davinci-1 GPU partition). In incompressible runs, STAR-CCM+ provided good weak scaling on up to 8 GPU nodes.

In three-way time-to-solution and scalability comparisons, PyFR at lowest order $p = 1$ outperformed Fluent by a factor of more than 8, and STAR-CCM+ by a factor of more than 4, in terms of wallclock time per degree of freedom. Notably, the coarse-grained PyFR run at $p = 4$ was also faster than the finite-volume simulations with Fluent and STAR-CCM+ in terms of wallclock per DoF. In evaluating the comparison, it is important to note that while considerable effort was put in homogenizing the parameter choices in Fluent, STAR-CCM+, and PyFR to achieve a level-playing field, differences due to, e.g., time integration methods and time step size could not be quantified.

In structural mechanics simulations of linear elasticity for the LPT blade with ABAQUS on davinci-1's CPU nodes, comparable figures were found for the speedup on up to 96 cores with the direct solver and iterative solver, but the iterative solver provided shorter time-to-solution and better usage of available memory. Regardless of the solver adopted, the accurate selection of the number of threads per MPI process through an internal solver parameter is paramount for achieving the best possible speedup when the resources available are limited.

The results reported in this paper open up a number of further avenues for investigation. First, the fidelity of simulations needs explicit validation, ideally both with experimental data and with simulations in the literature. The latter will require the use of inflow boundary conditions at an angle rather than purely horizontal. Second, PyFR results could be consolidated with runs at higher polynomial orders to fully exploit the accuracy and scalability potential of the implementation. This includes further coarse-graining the low order meshes to obtain more accurate simulations at similar computational costs [7], a feature not accessible to fixed-order methods in commercial packages. In addition, the mesh could be refined behind the trailing edge of the blade to improve the representation of the vortices in its wake. Third, the comparison between software frameworks could be extended to aircraft geometries, where scale- and boundary-layer-resolving simulations, and direct numerical simulations in particular, are bound to be computationally much more demanding. In terms of functionality, while Ansys Fluent and StarCCM+ already have capability for transonic simulations, further benchmarking on compressible aerodynamics tests will give the opportunity to test shock-capturing schemes in PyFR's high-order framework [8]. In addition, existing aerodynamics simulations chiefly concern flow over bluff-body fuselage, and implementation of moving meshes capability stands a great chance to provide higher-fidelity results in rotorcraft design, analysis, and simulation.

**CRediT authorship contribution statement**

**N. Sanguini:** conceptualization, investigation, methodology, software, validation, visualization, writing – original draft, writing – review & editing, data curation. **T. Benacchio:** conceptualization, investigation, methodology, software, validation, visualization, writing – original draft, writing – review & editing, project administration, supervision, data curation. **D. Malacrida:** investigation, software, validation, visualization, writing – review & editing. **F. Cipolletta:** conceptualization, investigation, methodology, software, validation, visualization, writing – review & editing, project administration, supervision. **F. Rondina:** conceptualization, investigation, methodology, software, validation, visualization, writing – original draft, writing – review & editing. **A. Sciarappa:** conceptualization, methodology, software, validation, writing – review & editing, supervision. **L. Capone:** conceptualization, project administration, supervision, funding acquisition.

**Acknowledgements**

This work greatly benefitted from useful discussions with PyFR's main developer Freddie D. Witherden and PyFR community forum pyfr.discourse.group. Furthermore, IT support by Alessandro Russo, Nicolò Magini and Manilo Trocano as well as useful feedback by Franco Ongaro, Carlo Cavazzoni, Michele Arra, Luigi Bottasso, Alessandro Scandroglio, Ivan Spisso, Gianni Bernini and Gregorio Frassoldati on an earlier version of the paper are gratefully acknowledged.